\documentclass[letterpaper]{article} 
\usepackage{aaai25}  
\usepackage{times}  
\usepackage{helvet}  
\usepackage{courier}  
\usepackage[hyphens]{url}  
\usepackage{graphicx} 
\usepackage{natbib}  
\usepackage{caption} 
\usepackage{algorithm}
\usepackage{algorithmic}

\usepackage{newfloat}
\usepackage{listings}
\DeclareCaptionStyle{ruled}{labelfont=normalfont,labelsep=colon,strut=off} 
\floatstyle{ruled}
\newfloat{listing}{tb}{lst}{}
\floatname{listing}{Listing}

\usepackage{color}
\usepackage{amsmath}
\usepackage{amssymb}
\usepackage{subfig}
\usepackage{booktabs}
\usepackage{tabularx}
\usepackage{multirow}
\usepackage{nameref}

\newcommand{\eat}[1]{}
\ifodd 1

\newcommand{\TODO}[1]{{\color{red}TODO:{#1}}}
\newcommand\beftext[1]{{\color[rgb]{0.5,0.5,0.5}{BEFORE:#1}}}
\else

\newcommand{\TODO}[1]{}
\newcommand{\beftext}[1]{}
\fi

\title{Large Language Model Enhanced Hard Sample Identification for Denoising Recommendation}
\author {
    Tianrui Song\textsuperscript{\rm 1},
    Wenshuo Chao\textsuperscript{\rm 1},
    Hao Liu\textsuperscript{\rm 1}\thanks{Corresponding author.}
}
\affiliations {
    \textsuperscript{\rm 1}The Hong Kong University of Science and Technology (Guangzhou)\\
    tsong847@connect.hkust-gz.edu.cn, wchao@connect.ust.hk, liuh@ust.hk
}
\usepackage{bibentry}

\begin{document}

\maketitle

\begin{abstract}
    Implicit feedback, often used to build recommender systems, unavoidably confronts noise due to factors such as misclicks and position bias.
    Previous studies have attempted to alleviate this by identifying noisy samples based on their diverged patterns, such as higher loss values, and mitigating the noise through sample dropping or reweighting.
    Despite the progress, we observe existing approaches struggle to distinguish hard samples and noise samples, as they often exhibit similar patterns, thereby limiting their effectiveness in denoising recommendations.
    To address this challenge, we propose a Large Language Model Enhanced Hard Sample Denoising (\textbf{LLMHD}) framework.
    Specifically, we construct an LLM-based scorer to evaluate the semantic consistency of items with the user preference, which is quantified based on summarized historical user interactions.
    The resulting scores are used to assess the hardness of samples for the pointwise or pairwise training objectives.
    To ensure efficiency, we introduce a variance-based sample pruning strategy to filter potential hard samples before scoring.
    Besides, we propose an iterative preference update module designed to continuously refine summarized user preference, which may be biased due to false-positive user-item interactions.
    Extensive experiments on three real-world datasets and four backbone recommenders demonstrate the effectiveness of our approach.
\end{abstract}

%

\section{Introduction}

Recommender systems are designed to learn user preferences and suggest items across various online platforms, such as e-commerce, news portals, and social networks \shortcite{He2020LightGCN, Luo2020SpatialOR, Zhang2023RLChargeIM}.
To train these systems, implicit feedback derived from user actions (e.g., clicks and purchases) is commonly employed due to its wide availability.
Typically, each observed interaction is assumed to reflect a user's genuine interest in an item and is therefore assigned a positive label, while non-interacted items are considered negative \shortcite{Ding2020Simplify, Wang2021DenoisingIF}. 
However, such a routine has recently been questioned that interacted items may be plagued by false-positive noise (e.g., due to misclicks or popularity bias), while non-interacted items may suffer from false-negative noise (e.g., due to position bias)~\shortcite{Wang2021ClicksCB}.
These noisy interactions lead to inaccurate estimation of user preferences, hindering the performance of recommendation systems.

Denoising recommendation has been proposed to mitigate the negative impact of noisy interactions through two primary strategies: 1) sample dropping and 2) sample re-weighting. 
Dropping methods aim to improve model performance by selecting clean samples and discarding noisy ones during training ~\shortcite{Wang2021DenoisingIF, Chen2021GeneralizedDW}. 
In contrast, re-weighting approaches assign lower weights to interactions identified as noisy, thereby reducing their influence on the model's learning process \shortcite{Wang2022EfficientBO, Gao2022SelfGuidedLT}. 
The success of these denoising techniques heavily depends on the accuracy of distinguishing between clean and noisy samples. 
Consequently, various data patterns have been explored as noisy signals ~\shortcite{Wang2021DenoisingIF, Gao2022SelfGuidedLT, Lin2023AutoDenoiseAD}.
To name a few,  loss value is one of the most commonly used signals, as noisy interactions typically exhibit higher loss values compared to clean ones \shortcite{Wang2021DenoisingIF, he2024double}.
In addition, other indicators such as prediction scores~\cite{Wang2022LearningRR} and gradients~\cite{Wang2022EfficientBO} have also been investigated to identify noisy samples.

\begin{figure}[t]
    \centering
    \includegraphics[width=0.9\linewidth]{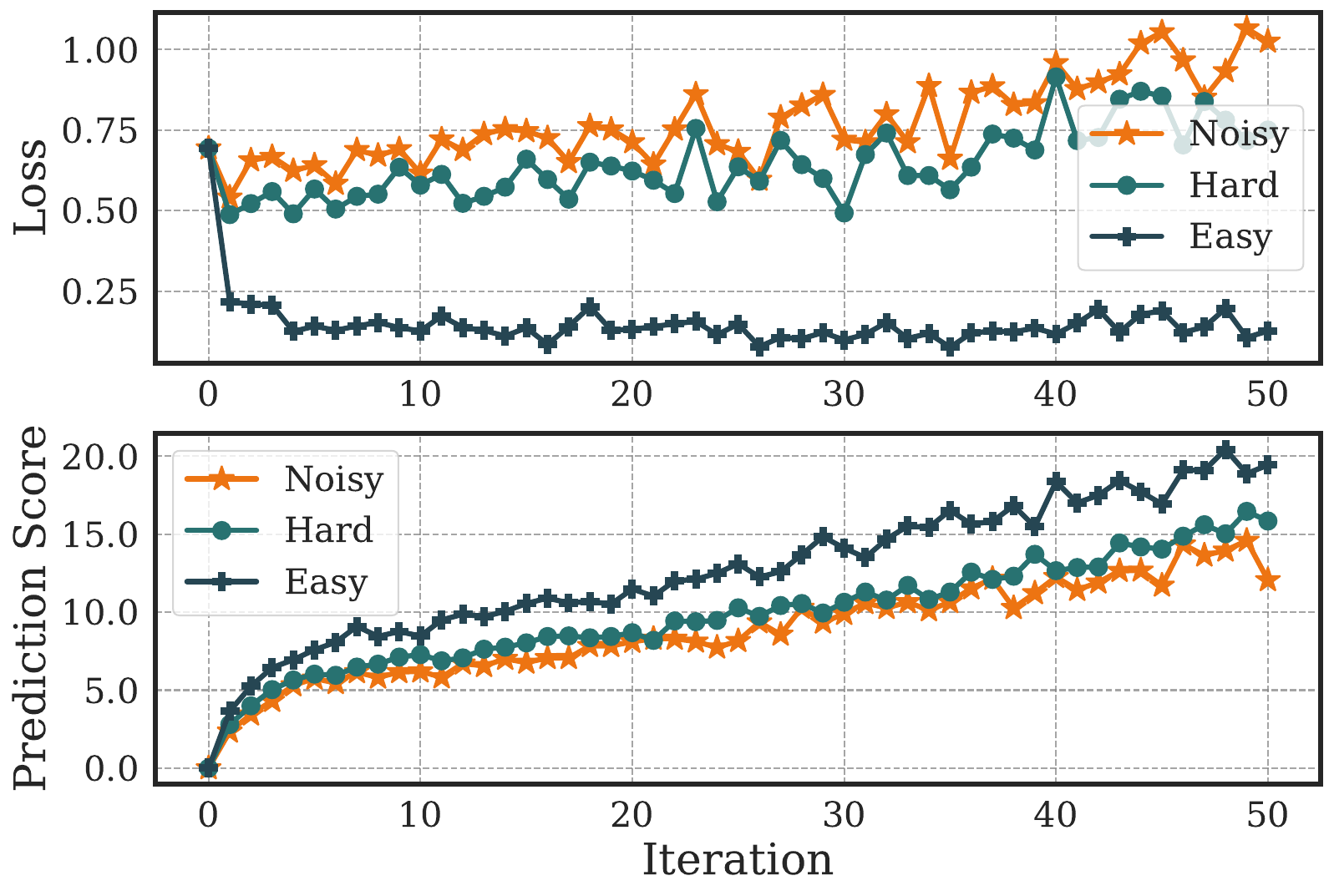}
    \caption{Loss values and prediction scores during training LightGCN on Yelp dataset. We observe that hard and noisy samples exhibit similar values in prediction score and loss, making it difficult to differentiate them.}
    \label{fig:Introduction}
\end{figure}

Despite significant advancements, existing methods often face the challenge of misidentifying hard samples as noisy ones.
As illustrated in Figure~\ref{fig:Introduction}, while noisy samples exhibit distinct patterns compared to easy samples, we observed that hard samples and noisy samples tend to present similar patterns in both prediction scores and loss values.
Consequently, previous denoising approaches that rely solely on data patterns \emph{struggle to accurately distinguish between hard and noisy samples}.
This misclassification is problematic because hard samples have been shown to be beneficial, both empirically~\shortcite{Gantner2012PersonalizedRF} and theoretically~\shortcite{Shi2023OnTT}.
Mistakenly treating hard samples as noise during the recommender training ultimately leads to suboptimal results.

Recently, Large Language Models (LLMs) have demonstrated a promising ability to understand user preferences \cite{Wu2023ASO} and enhance item semantics \cite{Wei2023LLMRec}, presenting a valuable opportunity to tackle the challenge of hard sample identification.
Our key insight is that LLMs can be harnessed to summarize user preference and act as a scorer to analyze the consistency between user preferences and items, thereby identifying hard samples with the resulting scores.
For example, when optimizing a model using a Bayesian Personalized Ranking (BPR) objective, 
the LLM scorer can effectively evaluate user preference scores of positive and negative items.
As a result, samples with similar positive and negative scores are pinpointed as hard samples because they are inherently incompatible with the BPR training objective, which aims to maximize the divergence in scores.
This allows us to mitigate the hard samples' misclassification issue in denoising recommender training.

However, leveraging LLMs for this task is nontrivial due to two primary challenges.
First, given the vast number of users and items, assessing the preferences of all users across all items is computationally intensive, especially considering the high inference cost of LLMs.
Second, while LLMs can derive user preference by concluding interacted items, the presence of false-positive items in historical interactions can lead to biased user preference summarization.

To address the challenges mentioned above, we propose a \textbf{L}arge \textbf{L}anguage \textbf{M}odel Enhanced \textbf{H}ard Sample \textbf{D}enoising (\textbf{LLMHD}) framework for recommendation, which comprises three key modules: Variance-based Sample Pruning, LLM-based Sample Scoring, and Iterative Preference Updating. 
To ensure efficiency, we first introduce a variance-based pruning strategy that progressively selects a small subset of hard sample candidates.
Following this, we construct the LLM-based Sample Scoring module, where hard samples are identified by evaluating how well they satisfy the training objective.
Specifically, the LLM scores the user preference for a given item by summarizing user preference, assesses the sample's hardness based on the pointwise or pairwise training objective, and determines whether it qualifies as a hard sample.
Additionally, to enhance the accuracy of the summarized user preferences, we propose an Iterative Preference Updating module.
It refines user preferences by adjusting for items that are mistakenly identified or overlooked during the summarization process, thereby improving the overall reliability of the LLMHD framework.

 Our main contributions are summarized as follows:
\begin{itemize} 
    \item We propose \textbf{LLMHD}
    , a novel denoising recommendation approach that differentiates between hard and noisy samples leveraging LLMs. To the best of our knowledge, this is the first attempt to integrate LLMs into denoising recommendations. 
    \item The LLMHD addresses efficiency concerns through a variance-based sample pruning process. Furthermore, we enhance the effectiveness of the model by employing an iterative preference updating strategy, improving the LLMs' understanding of genuine user preferences. 
    \item Extensive experiments conducted on three real-world datasets and four backbone recommenders demonstrate the effectiveness of our method. The results show that LLMHD delivers impressive performance and robust noise resilience. 
\end{itemize}

\section{Related Work}

\subsection{Denoise Recommendation}

Recommenders are pointed out to be affected by users’ unconscious behaviors \shortcite{Wang2021ClicksCB}, leading to noisy data.
As a result, many efforts are dedicated to alleviating the problem.
These approaches can be categorized into two paradigms: sample dropping \shortcite{Gantner2012PersonalizedRF, Lin2023AutoDenoiseAD} and sample re-weighting \shortcite{Wang2022EfficientBO, Gao2022SelfGuidedLT}.
Sample dropping methods aim to keep clean samples and discard noisy ones.
For instance, T-CE \shortcite{Wang2021DenoisingIF} observes that noisy samples exhibit high loss values and remove them during training.
IR \shortcite{Wang2021ImplicitFA} iteratively generates pseudo-labels to discover noisy examples.
Sample re-weighting methods try to mitigate the impact of noisy samples by assigning lower weights to them.
Typically, R-CE \shortcite{Wang2021DenoisingIF} assigns lower weights to noisy samples according to the prediction score.
BOD \shortcite{Wang2022EfficientBO} considers the weight assignment as a bi-level optimization problem.
Although these methods achieve promising results, they rely on data patterns to recognize noisy samples (e.g., loss values, and prediction scores) leading to difficulties in identifying hard samples from noise samples as they exhibit similar patterns.

\subsection{LLMs for Recommendation}

Large Language Models (LLMs) are effective tools for the Natural Language Processing field and have gained significant attention in the domain of Recommendation Systems (RS).
For the adaption of LLMs in recommendations, existing works can be divided into three categories \shortcite{Wu2023ASO}: LLM as RS, LLM Embedding for RS, and LLM token for RS.
The LLM as RS aims to transform LLMs into effective recommendation systems \cite{chao2024makelargelanguagemodel}, such as LC-Rec \shortcite{Zheng2023AdaptingLL} and LLM-TRSR \shortcite{Zheng2024HarnessingLL}.
In contrast, the LLM embedding and LLM token for RS views the language model as an enhancer, where embeddings and tokens generated by LLMs are utilized for promoting recommender systems.
The former typically adopts embeddings related to users and items, incorporating semantic information in the recommender \cite{Ren2023RepresentationLW}.
While the latter generates text tokens to capture potential preferences through user and item semantics \cite{Wei2023LLMRec, Xi2023TowardsOR}.
Despite the progress, these methods overlook the potential of LLMs in enhancing data denoising for recommendation.

\section{Preliminary} 
The objective of training a recommender system is to learn a scoring function $\hat{y}_{u,i} = f_{\theta}(u,i)$ from interactions between users $u \in \mathcal{U}$ and items $i \in \mathcal{I}$. 
We assume that user-interacted items ${y}^{*}_{ui}= 1$ are preferred by the user, while those not interacted $y^{*}_{ui}= 0$ are not.
To optimize the scoring function $f_{\theta}(u,i)$, We employ Bayesian Personalized Ranking (BPR) loss and Binary Cross-Entropy (BCE) loss as loss function $\mathcal{L}_{rec}$. These are formulated as:
\begin{scriptsize}
\begin{equation}
    \label{eq:bpr}
    \mathcal{L}_{BPR}(\mathcal{D}^{*}) = -\mathop{\mathbb{E}}\limits_{(u,i,j) \sim \mathbf{P}_{\mathcal{D}^{*}}} log(\sigma(\hat{y}_{u,i} - \hat{y}_{u,j})),
\end{equation}
\end{scriptsize}
\begin{scriptsize}
\begin{equation}
    \label{eq:bce}
    \mathcal{L}_{BCE}(\mathcal{D}^{*}) = -\mathop{\mathbb{E}}\limits_{(u,i,y_{ui}^{*}) \sim \mathbf{P}_{\mathcal{D}^{*}}} y_{ui}^{*}log(\hat{y}_{u,i}) + (1-y^{*}_{ui})log(1-\hat{y}_{u,i}),
\end{equation}
\end{scriptsize}
where $j$ denotes sampled negative items according to the pairwise sampling distribution $\mathbf{P}_{\mathcal{D}^{*}}$, and $\mathcal{D}^{*}= \{(u,i, y_{ui}^{*}) \mid u \in \mathcal{U}, i \in \mathcal{I}\}$ represents the interaction dataset. The optimal parameter set $\theta^{*}$ is obtained by minimizing the loss function:
\begin{align}
    \label{eq:recobj}
    \theta^{*} = \mathop{argmin}\limits_{\theta}\mathcal{L}_{rec}(\mathcal{D}^{*}),
\end{align}
But this assumption is unreliable for two reasons: 
\emph{(1) False positive issue}, user-interacted items might not reflect real user preference due to factors such as accidental clicks and position bias.
\emph{(2) False negative issue}, non-interacted items are not necessarily user dislikes, they may have been overlooked due to factors such as suboptimal display positions.
These issues introduce noisy interactions, formally defined as $\tilde{\mathcal{D}} = \{(u,i,\tilde{y}) \mid \tilde{y} \neq y^{*}\}$. To address this, we formulate the \emph{denoising recommender training task} as:
\begin{equation}
    \label{eq:noiseobj}
    \theta^{*} = \mathop{argmin}\limits_{\theta}\mathcal{L}_{rec}(\mathcal{D}^{*} \cup \tilde{\mathcal{D}}),
\end{equation}
aiming to learn high-quality recommender with parameters $\theta^{*}$ by eliminating the effect of noisy samples.
In this work, we focus on the challenge of hard samples, which are often mistakenly identified as noisy samples in existing denoising approaches, leading to suboptimal performance.

\section{Proposed Method}
\begin{figure*}[htbp!]
  \centering
  \includegraphics[width=\textwidth]{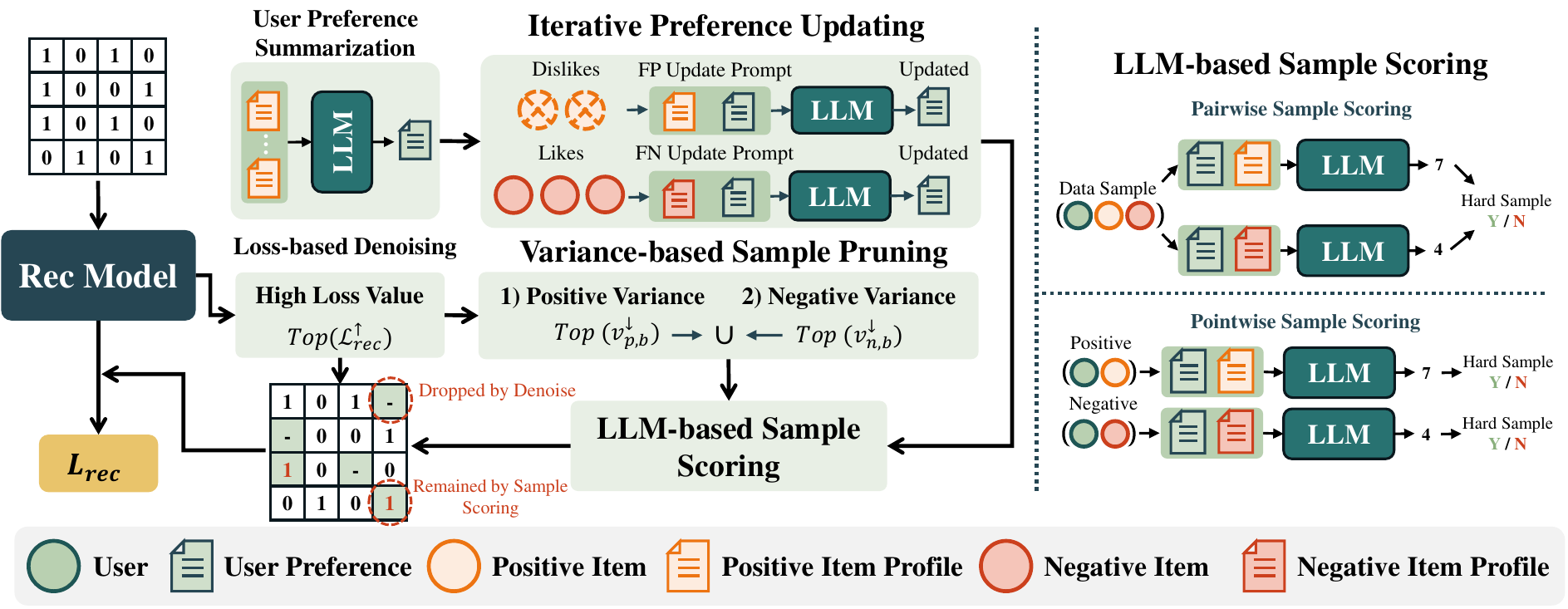}
  \caption{
        The overview of the LLMHD framework.
        LLMHD leverages LLMs to differentiate hard and noisy samples, thereby enhancing the denoising recommender training task.
        The framework identifies hard samples through three main modules: \textit{(1) Variance-based Sample Pruning}, \textit{(2) LLM-based Sample Scoring}, and \textit{(3) Iterative Preference Updating}.
    }
  \label{fig:overview}
\end{figure*}

To differentiate hard and noisy samples when denoising, we proposed the LLMHD framework, as illustrated in Figure \ref{fig:overview}. 
Before diving into the details of each module, we assume that each item $i$ is accompanied by a text profile $\mathcal{P}_{i}$.
Additionally, we summarize the user's preference ${\mathcal{P}_{u} = LLMs(T_{sum}(\{\mathcal{P}_{i} \mid y_{u,i}=1\}))}$ by the profiles of interacted items with a prompt template $T_{sum}$ designed for LLMs.
Our LLMHD identifies hard samples through three key modules:
(1) Variance-based Sample Pruning, (2) LLM-based Sample Scoring, and (3) Iterative Preference Updating.
Variance-based Sample Pruning reduces the computation of calling LLMs by selecting a subset of hard sample candidates.
LLM-based Sample Scoring evaluates the hardness of samples based on user preferences.
Iterative Preference Updating refines the understanding of user preference, ensuring accurate identification of hard samples.

\subsection{Loss-based Denoising} 
We first introduce the denoising module implemented based on the widely accepted assumption \shortcite{Wang2021DenoisingIF} that samples with higher loss values are more likely to be noisy.
Specifically, for each data sample $b$ in the mini-batch $\mathcal{B}$, we calculate the corresponding loss value $l(b)$ and sort all samples in the ascending order,
\begin{equation}
    l(b_{1})^{\uparrow} < l(b_{2})^{\uparrow} < \cdots < l(b_{|\mathcal{B}|})^{\uparrow}, \quad b_j \in \mathcal{B},
\end{equation}
where $|\mathcal{B}|$ denotes the batch size.
This operation assists the noisy sample identification, which we formulate as $\mathcal{B}_{N}$,
\begin{equation}
    \label{eq:basedenoise}
    \mathcal{B}_{N} = \left\{b_j \mid l(b_{j})^{\uparrow} \geq l(b_{\varepsilon_{l}})^{\uparrow}\right\},
\end{equation}
where $T$ denotes the current training iteration. The $\varepsilon_{l}$ represents a dynamic threshold, calculated as,
\begin{equation}
    \label{eq:dt}
    \varepsilon_{l} = \min(\frac{1}{\alpha}T, \varepsilon_{l}^{\max}|\mathcal{B}|),
\end{equation}
where $\varepsilon_{l}^{\max}$ is a hyper-parameter representing the maximum noise ratio, and $\alpha$ is a factor that modulates the growth rate of the noise threshold. 
The $\varepsilon_{l}$ increases as the stability of prediction scores incrementally improves during training, following previous works \cite{Wang2021DenoisingIF}.
It is worth mentioning that the $\mathcal{B}_{N}$ inadvertently contain hard samples, given that both hard and noisy samples manifest similar patterns in loss values.
This requires further refinement to distinguish genuine noisy data and hard samples.

\subsection{Variance-based Sample Pruning \label{sec:vss}}

Although it is possible to present all identified noisy samples $\mathcal{B}_{N}$ to the LLMs for scoring, this approach would be prohibitively time-consuming due to the massive interactions in the recommender system.
Specifically, hard sample candidates are selected based on the observation of previous work ~\shortcite{Ding2020Simplify}, which demonstrated that hard samples exhibit relatively higher prediction score variance compared to noisy samples.
Therefore, for samples $b \in \mathcal{B}_{N}$, we calculate the prediction scores variance of positive $v_{p,b}$ and negative $v_{n,b}$ items across multiple epochs (see Equation \ref{eq:variance}).
Then sort them in descending order based on $v_{p}$ and $v_{n}$ respectively,
\begin{equation}
    v_{p,1}^{\downarrow} > v_{p,2}^{\downarrow} > \cdots > v_{p,b}^{\downarrow} > \cdots > v_{p, |\mathcal{B}_{N}^{p}|}^{\downarrow}, \space b \in \mathcal{B}_{N},
\end{equation}
\begin{equation}
    v_{n,1}^{\downarrow} > v_{n, 2}^{\downarrow} > \cdots > v_{n,b}^{\downarrow} > \cdots > v_{n, |\mathcal{B}_{N}^{n}|}^{\downarrow}, \space b \in \mathcal{B}_{N},
\end{equation}
where $|\mathcal{B}_{N}^{p}|$ and $|\mathcal{B}_{N}^{n}|$ denotes the number of positive and negative items in the $\mathcal{B}_{N}$ respectively. Hard sample candidates $\mathcal{B}_{HC}$ are collected by,
\begin{equation}
    \small
    \mathcal{B}_{HC} = \{b_{j}|v_{p, j}^{\downarrow} \geq v_{p, \varepsilon_{v} |\mathcal{B}_{N}^{p}|}^{\downarrow}\} \cup \{b_{j}|v_{n, b_{j}}^{\downarrow} \geq v_{n, \varepsilon_{v} |\mathcal{B}_{N}^{n}|}^{\downarrow}\},
\end{equation}
where $\varepsilon_{v}\in[0,1]$ denotes the proportion of hard samples.
With the increasing $|\mathcal{B}_{N}|$ more candidates will be selected in latter training iterations and provided to LLM-based Sample Scoring to identify hard samples further.

\subsection{LLM-based Sample Scoring \label{sec:llmps}}

Owing to the resemblance in data patterns between hard and noisy samples, distinguishing them solely through numerical disparities is ineffective.
To eliminate this issue, we introduce the LLM-based Sample Scoring method.
LLMs act as scorer to provide auxiliary information that evaluates the sample's hardness.
Formally, we prompt LLMs to score the user preference for item $s_{u,i}$ with a template $T_{score}(\mathcal{P}_{u}, \mathcal{P}_{i})$ that wraps the user preference text $\mathcal{P}_{u}$ and item profile $\mathcal{P}_{i}$,
\begin{equation}
    \small
    \label{eq:uis}
    s_{u,i} = LLMs(T_{score}(\mathcal{P}_{u},\mathcal{P}_{i})).
\end{equation}
The resulting score $ s_{u,i}$ is adopted to specify the sample's hardness by analyzing its compatibility with the training objective.
Lower compatibility samples are considered harder as they are more challenging to satisfy the objective.
Given that most recommenders are trained to minimize pointwise (e.g., BCE) or pairwise (e.g., BPR) losses, we devise two paradigms for hard sample identification: (1) Pointwise Preference Scoring, and (2) Pairwise Preference Scoring.

\subsubsection{Pointwise Sample Scoring}
The pointwise BCE loss, as shown in Equation \ref{eq:bce}, aims at reducing the classification uncertainty of a (user, item) pair.
For a data sample $(u, i_{pos})$ or $(u, i_{neg})$, if the user's preference for the item is ambiguous, the sample is of low compatibility with the training objective.
Therefore, positive pair with lower $s_{u,i_{pos}}$ and negative pair with higher $s_{u,i_{neg}}$ are harder samples, thereby hard samples are identified by,
\begin{equation}
    \label{eq:hardIdentifypa}
    \small
    \mathbb{I}_{\text{point}}(u,i) = \left\{
            \begin{aligned}
                1, & & \text { if } s_{u,i} < \varepsilon_{pos} \text{ and } y_{u,i}=1\\
                1, & & \text { if } s_{i,i} > \varepsilon_{neg} \text{ and } y_{u,i}=0\\
                0, & & \text { otherwise,}
            \end{aligned}
        \right.
\end{equation}
where the $\varepsilon_{pos}$ and $\varepsilon_{neg}$ are thresholds that control the hardness.
In addition, since previous works \shortcite{Wang2021DenoisingIF} discussed that fitting harder samples at the early training stage might hurt the generalization ability,  
we smoothly change $\varepsilon_{pos}$ and $\varepsilon_{neg}$ during each training iteration $T$ as follows,
\begin{equation}
    \small
    \varepsilon_{pos}=max(\varepsilon_{pos}^{max}-\frac{1}{\alpha}T, \varepsilon_{pos}^{min}),
\end{equation}
\begin{equation}
    \small
    \varepsilon_{neg}=min(\varepsilon_{neg}^{min}+\frac{1}{\alpha}T, \varepsilon_{neg}^{max}),
\end{equation}
where $\varepsilon_{pos}^{max}, \varepsilon_{pos}^{min}, \varepsilon_{neg}^{max}, \varepsilon_{neg}^{min}$ are hyper-parameters.
In this way, harder positive $(u, i_{pos})$ and negative $(u, i_{neg})$ samples will be identified in the latter iterations, benefiting the recommender by gradually increasing the hardness.

\subsubsection{Pairwise Sample Scoring}
Similar to the pointwise sample scoring, we identify hard samples under the pairwise training schema.
Specifically, according to Equation \ref{eq:bpr}, the pairwise BPR loss aims to maximize the divergence of prediction scores between positive and negative items.
For a sample $(u, i_{p}, i_{n})$, if the user's preference for the positive item does not significantly surpass that for the negative, the sample is less compatible with the objective.
Therefore, hard samples are identified through the indicator function, 
\begin{equation}
    \label{eq:hardIdentify}
    \mathbb{I}_{pair}(s_{u,i_{p}}-s_{u,i_{n}} > \varepsilon_{pair}),
\end{equation}
where the threshold $\varepsilon_{pair}$ also gradually decreases to increase the hardness by the number of iteration $T$,
\begin{equation}
    \varepsilon_{pair} = max(\varepsilon_{pair}^{max}-\frac{1}{\alpha}T, \varepsilon_{pair}^{min}).
\end{equation}
Based on the above technique, we differentiate hard samples in both pointwise and pairwise training schema.

\subsection{Iterative Preference Updating \label{sec:ipu}}

Accurate user preference $\mathcal{P}{u}$ is critical for effective LLM sample scoring.
However, the $\mathcal{P}{u}$ summarized based on interacted items do not fully capture user interests due to the inclusion of disliked items, i.e., false-positives, and the exclusion of liked items, i.e., false-negatives.
To mitigate this problem, we refine user preferences iteratively by excluding dislikes and incorporating likes.
For every epoch $t$, we calculate the variance score $v_{d}$ of user-item pairs $d=(u,i)$,
\begin{equation}
    \label{eq:variance}
    \small
    v_{d} = \frac{1}{m}\sum_{j=t-m+1}^{t}\left (\hat{y}_d^{j} - \frac{\sum_{j=t-m+1}^{t}\hat{y}_d^{j}}{m}\right )^{2},
\end{equation}
where $\hat{y}_d^{j}$ is the prediction score of user-item pair $d$ in the $j$-th training epoch, and the variance $v_{d}$ is calculated over $m$ time intervals prior to the $t$-th training iteration.
We divided variance scores into two groups, positive and negative samples, and ordered from lowest to highest,
\begin{equation}
    v_{d^{\text{p}}_{1}}^{\uparrow} < v_{d^{\text{p}}_{2}}^{\uparrow} < \cdots < v_{d^{\text{p}}_{|\mathcal{D}_{pos}|}}^{\uparrow}, d^{\text{p}}_{k} \in \mathcal{D}_{pos},
\end{equation}
\begin{equation}
    v_{d^{\text{n}}_{1}}^{\uparrow} < v_{d^{\text{n}}_{2}}^{\uparrow} < \cdots < v_{d^{\text{n}}_{|\mathcal{D}_{neg}|}}^{\uparrow}, d^{\text{n}}_{k} \in \mathcal{D}_{neg},
\end{equation}
where $d^{\text{p}}_{k}, d^{\text{n}}_{k}$ are the $k$-th positive and negative sample respectively.
To identify whether a sample is a false positive or false negative in the $j$-th epoch,
we use the indicators $\mathbb{I}_{\text{fp}}^{j}(d_{k}^{\text{p}} \leq d_{\varepsilon{l}}^{\text{p}})$ and $\mathbb{I}_{\text{fn}}^{j}(d_{k}^{\text{n}} \geq d_{\varepsilon_{l}}^{\text{n}})$ respectively.
The threshold $\varepsilon_{l}$ employed here follows the same definition as introduced in Equation \ref{eq:dt}.
We design a robust mechanism to select confident items for preference updates.
Formalized as follows,
\begin{equation}
    \small
    \mathbb{I}_{\text{FP}}(\sum_{j=0}^{t} \mathbb I(d_{k}^{\text{p}}) \geq \varepsilon_{\gamma}), \quad
    \mathbb{I}_{\text{FN}}(\sum_{j=0}^{t} \mathbb I(d_{k}^{\text{n}}) \geq \varepsilon_{\gamma}),    
\end{equation}
the $\varepsilon_{\gamma}$ is a confidence threshold.
We then leverage LLMs to refine preference $\mathcal{P}_{u}$ based on identified false-positives $(u,i_{\text{fp}})$ and false negatives $(u,i_{\text{fp}})$ with the template $T_{\text{FP}}(\mathcal{P}_{u}, \mathcal{P}_{i_{\text{fp}}})$ and $T_{\text{FN}}(\mathcal{P}_{u}, \mathcal{P}_{i_{\text{fn}}})$,
\begin{equation}
    \mathcal{P}_{u}^{*} = LLM(T_{\text{FP}}(\mathcal{P}_{u}, \mathcal{P}_{i_{\text{fp}}})),
\end{equation}
\begin{equation}
    \mathcal{P}_{u}^{*} = LLM(T_{\text{FN}}(\mathcal{P}_{u}, \mathcal{P}_{i_{\text{fn}}})),
\end{equation}
where $\mathcal{P}_{u}^{*}$ is the updated user preference text description.
The template $T_{\text{FP}}$ intend to add descriptioins about $i_{\text{fp}}$ in the user preference $\mathcal{P}_{u}$, while the $T_{\text{FN}}$ reduce the feature of $i_{\text{fn}}$.

\subsection{Denoising Training with Hard Samples}
The denoising training is done by keeping hard samples and dropping noisy samples.
We first define the set of identified hard samples $\mathcal{B}_{H}$ as,
\begin{equation}
    \mathcal{B}_{H} = \{b_{j} \mid \mathbb{I}_{\text{LLM}}(b_{j})=1, b_{j} \in \mathcal{B}_{HC}\},
\end{equation}
where the $\mathbb{I}_{\text{LLM}}$ is either $\mathbb{I}_{point}$ or $\mathbb{I}_{pair}$ based on the format of data samples.
The recommendation loss $\mathcal{L}_{rec}$ is then calculated in the following format,
\begin{equation}
    \mathcal{L}_{rec}((\mathcal{B} \setminus \mathcal{B}_{N}) \cup \mathcal{B}_{H}).
\end{equation}
In this way, hard samples have remained and the noisy samples are dropped while training the recommender.

\section{Experiments}
We compare LLMHD with state-of-the-art denoise approaches on four backbones and three real-world datasets to demonstrate the effectiveness of our method.
Experiments are directed by the following research questions (RQs):
\begin{itemize}
\item \textbf{RQ1:} How does LLMHD performs compared with other state-of-the-art denoise baselines across the datasets?
\item \textbf{RQ2:} Does the LLMHD demonstrate robustness when tackling different levels of noisy data?
\item \textbf{RQ3:} What is the effect of different components and hyper-parameters within the LLMHD on performance?
\end{itemize}

\begin{table*}[t]
    \small
    \centering
    \setlength{\tabcolsep}{4.4pt}
    \begin{tabularx}{\textwidth}{c|c|cccc|cccc|cccc}
        \toprule
        \multicolumn{2}{c|}{\textbf{Dataset}} & \multicolumn{4}{c|}{\textbf{Amazon-book}} & \multicolumn{4}{c|}{\textbf{Yelp}} & \multicolumn{4}{c}{\textbf{Steam}}\\
        \midrule
        \textbf{Backbone} & \textbf{Method} & \textbf{R@5} & \textbf{R@10} & \textbf{N@5} & \textbf{N@10} & \textbf{R@5} & \textbf{R@10} & \textbf{N@5} & \textbf{N@10} & \textbf{R@5} & \textbf{R@10} & \textbf{N@5} & \textbf{N@10} \\
        \hline\hline
        
        \multirow{8}*{NGCF} & BCE & 0.0353 & 0.0570 & 0.0365 & 0.0438 & 0.0236 & 0.0431 & 0.0283 & 0.0350 & 0.0223 & 0.0405 & 0.0236 & 0.0305 \\
        ~ & BPR & 0.0389 & 0.0651 & 0.0406 & 0.0494 & 0.0280 & 0.0495 & 0.0338 & 0.0405 & 0.0381 & 0.0629 & 0.0453 & 0.0525 \\
        ~ & T-CE & 0.0393 & 0.0650 & 0.0402 & 0.0489 & 0.0259 & 0.0450 & 0.0313 & 0.0373 & 0.0257 & 0.0448 & 0.0288 & 0.0354 \\
        ~ & R-CE & 0.0366 & 0.0587 & 0.0369 & 0.0444 & 0.0254 & 0.0438 & 0.0302 & 0.0360 & 0.0236 & 0.0435 & 0.0254 & 0.0328 \\
        ~ & RGCF & \underline{0.0415} & 0.0658 & \underline{0.0422} & 0.0502 & \underline{0.0287} & 0.0485 & 0.0344 & 0.0406 & \underline{0.0401} & \underline{0.0644} & \underline{0.0472} & \underline{0.0543} \\
        ~ & DCF & 0.0398 & 0.0617 & 0.0399 & 0.0472 & 0.0281 & \underline{0.0488} & \underline{0.0353} & \underline{0.0414} & 0.0264 & 0.0446 & 0.0308 & 0.0365 \\
        \cline{2-14}
        \rule{0pt}{9pt} ~ & \textbf{$\text{LLMHD}_{\text{BCE}}$} & 0.0406 & \underline{0.0668} & 0.0413 & \underline{0.0503} & 0.0276 & 0.0477 & 0.0329 & 0.0386 & 0.0267 & 0.0459 & 0.0297 & 0.0364 \\
        ~ & \textbf{$\text{LLMHD}_{\text{BPR}}$} & \textbf{0.0455} & \textbf{0.0743} & \textbf{0.0455} & \textbf{0.0552} & \textbf{0.0338} & \textbf{0.0579} & \textbf{0.0398} & \textbf{0.0474} & \textbf{0.0418} & \textbf{0.0696} & \textbf{0.0496} & \textbf{0.0579} \\
        
        \midrule 
        \multirow{8}*{LightGCN} & BCE & 0.0558 & 0.0849 & 0.0565 & 0.0665 & 0.0390 & 0.0660 & 0.0481 & 0.0557 & 0.0448 & 0.0732 & 0.0529 & 0.0612 \\
        ~ & BPR & 0.0587 & 0.0904 & 0.0598 & 0.0704 & 0.0359 & 0.0609 & 0.0446 & 0.0516 & 0.0510 & 0.0828 & 0.0597 & 0.0693 \\
        ~ & T-CE & 0.0590 & 0.0895 & 0.0592 & 0.0697 & 0.0401 & 0.0677 & 0.0504 & 0.0580 & 0.0463 & 0.0758 & 0.0555 & 0.0640 \\
        ~ & R-CE & 0.0557 & 0.0834 & 0.0566 & 0.0658 & 0.0389 & 0.0650 & 0.0474 & 0.0550 & 0.0461 & 0.0757 & 0.0543 & 0.0630 \\
        ~ & RGCF & \underline{0.0619} & \underline{0.0956} & \underline{0.0644} & \underline{0.0753} & \underline{0.0420} & \underline{0.0693} & 0.0501 & 0.0579 & \underline{0.0519} & \underline{0.0849} & \underline{0.0599} & \underline{0.0702} \\
        ~ & DCF & 0.0590 & 0.0898 & 0.0596 & 0.0701 & 0.0403 & 0.0680 & 0.0503 & 0.0579 & 0.0477 & 0.0778 & 0.0562 & 0.0650 \\
        \cline{2-14}
        \rule{0pt}{9pt} ~ & \textbf{$\text{LLMHD}_{\text{BCE}}$} & 0.0607 & 0.0921 & 0.0607 & 0.0711 & 0.0408 & 0.0689 & \underline{0.0514} & \underline{0.0589} & 0.0469 & 0.0767 & 0.0563 & 0.0647 \\
        ~ & \textbf{$\text{LLMHD}_{\text{BPR}}$} & \textbf{0.0652} & \textbf{0.0999} & \textbf{0.0655} & \textbf{0.0767} & \textbf{0.0427} & \textbf{0.0731} & \textbf{0.0518} & \textbf{0.0611} & \textbf{0.0536} & \textbf{0.0867} & \textbf{0.0624} & \textbf{0.0722} \\

        \midrule 
         \multirow{8}*{SGL} & BCE & 0.0589 & 0.0902 & 0.0604 & 0;.0707 & 0.0377 & 0.0655 & 0.0470 & 0.0548 & 0.0433 & 0.0682 & 0.0505 & 0.0676 \\
        ~ & BPR & 0.0608 & 0.0956 & 0.0621 & 0.0736 & 0.0373 & 0.0629 & 0.0465 & 0.0538 & 0.0529 & 0.0838 & 0.0613 & 0.0704 \\
        ~ & T-CE & 0.0602 & 0.0909 & 0.0622 & 0.0720 & 0.0408 & 0.0697 & 0.0502 & 0.0587 & 0.0449 & 0.0720 & 0.0532 & 0.0609 \\
        ~ & R-CE & 0.0591 & 0.0901 & 0.0601 & 0.0702 & 0.0386 & 0.0645 & 0.0476 & 0.0550 & 0.0456 & 0.0732 & 0.0538 & 0.0618 \\
        ~ & RGCF & \underline{0.0675} & \underline{0.1049} & \underline{0.0681} & \underline{0.0808} & \underline{0.0416} & \underline{0.0715} & \underline{0.0512} & \underline{0.0606} & \textbf{0.0552} & \underline{0.0881} & \underline{0.0639} & \underline{0.0736} \\
        ~ & DCF & 0.0626 & 0.0933 & 0.0641 & 0.0740 & 0.0413 & 0.0683 & 0.0506 & 0.0583 & 0.0455 & 0.0727 & 0.0536 & 0.0615 \\
        \cline{2-14}
        \rule{0pt}{9pt} ~ & \textbf{$\text{LLMHD}_{\text{BCE}}$} & 0.0615 & 0.0931 & 0.0640 & 0.0739 & 0.0414 & 0.0708 & 0.0509 & 0.0596 & 0.0462 & 0.0742 & 0.0543 & 0.0619 \\
        ~ & \textbf{$\text{LLMHD}_{\text{BPR}}$} & \textbf{0.0693} & \textbf{0.1051} & \textbf{0.0717} & \textbf{0.0837} & \textbf{0.0426} & \textbf{0.0718} & \textbf{0.0523} & \textbf{0.0619} & \underline{0.0546} & \textbf{0.0887} & \textbf{0.0641} & \textbf{0.0739} \\

        \midrule
        \multirow{8}*{NCL} & BCE & 0.0574 & 0.0871 & 0.0598 & 0.0694 & 0.0391 & 0.0647 & 0.0477 & 0.0548 & 0.0450 & 0.0731 & 0.0529 & 0.0612 \\
        ~ & BPR & 0.0605 & 0.0942 & 0.0628 & 0.0740 & 0.0369 & 0.0609 & 0.0451 & 0.0515 & 0.0511 & 0.0835 & 0.0602 & 0.0698 \\
        ~ & T-CE & 0.0599 & 0.0898 & 0.0619 & 0.0719 & 0.0411 & 0.0679 & 0.0507 & 0.0582 & 0.0461 & 0.0751 & 0.0543 & 0.0627 \\
        ~ & R-CE & 0.0585 & 0.0874 & 0.0604 & 0.0696 & 0.0399 & 0.0655 & 0.0487 & 0.0558 & 0.0459 & 0.0750 & 0.0540 & 0.0625 \\
        ~ & RGCF & \underline{0.0694} & \underline{0.1045} & \underline{0.0706} & \underline{0.0819} & 0.0396 & 0.0660 & 0.0480 & 0.0560 & \underline{0.0534} & \textbf{0.0863} & \underline{0.0621} & \textbf{0.0718} \\
        ~ & DCF & 0.0619 & 0.0929 & 0.0624 & 0.0727 & \underline{0.0424} & \underline{0.0696} & 0.0513 & 0.0589 & 0.0465 & 0.0759 & 0.0550 & 0.0635 \\
        \cline{2-14}
        \rule{0pt}{9pt} ~ & \textbf{$\text{LLMHD}_{\text{BCE}}$} & 0.0609 & 0.0915 & 0.0629 & 0.0730 & 0.0417 & 0.0690 & \underline{0.0517} & \underline{0.0591} & 0.0468 & 0.0762 & 0.0551 & 0.0633 \\
        ~ & \textbf{$\text{LLMHD}_{\text{BPR}}$} & \textbf{0.0719} & \textbf{0.1053} & \textbf{0.0741} & \textbf{0.0846} & \textbf{0.0432} & \textbf{0.0716} & \textbf{0.0542} & \textbf{0.0620} & \textbf{0.0540} & \underline{0.0861} & \textbf{0.0624} & \underline{0.0717} \\
        \bottomrule
    \end{tabularx}
    \caption{
        Performance comparison of backbone recommenders trained with different denoising approaches.
        R and N refer to Recall and NDCG, respectively.
        The highest scores are in \textbf{bold}, and the runner-ups are with \underline{underline}.
        All results are statistically significant according to the t-tests with a significance level of $p < 0.01$.
    }
   \label{table:performance}
\end{table*}

\subsection{Experiment Settings}

\subsubsection{Datasets.}

We conduct evaluations of our LLMHD on three public datasets:
(1) \textbf{Amazon-Books} collected from the Amazon platform. We conduct experiments on the book subcategories.
(2) \textbf{Yelp} is a large-scale dataset that provides check-in history.
(3) \textbf{Steam} consists of users and electronic games on the Steam platform.
Since we adopt the item profile provided in \cite{Ren2023RepresentationLW}, we process datasets following their settings.
Details are in the Appendix.

\subsubsection{Evaluation Metrics.}
Following existing works on denoising recommendation \shortcite{Wang2021ImplicitFA, he2024double}, we report the results w.r.t. two widely used metrics: NDCG@\textit{K} and Recall@\textit{K}, where higher scores indicate better performance.
For a comprehensive comparison, we set the K as 5 and 10 for both metrics on all three datasets.

\subsubsection{Baselines.}
To evaluate the performance of LLMHD, we apply it to the following backbone recommenders:
\begin{itemize}
    \item \textbf{NGCF} \shortcite{Wang2019NeuralGC} models the user-item interaction graph with GNN for collaborative filtering.
    \item \textbf{LightGCN} \shortcite{He2020LightGCN} removes the feature transformation and non-linear activation in NGCF.
    \item \textbf{SGL} \shortcite{Wu2020SelfsupervisedGL} generates positive views of with model-level node and edge dropout for self-supervised learning.
    \item \textbf{NCL} \shortcite{Lin2022ImprovingGC} exploit the neighborhood structure to conduct self-supervised learning in graph recommenders. 
\end{itemize}
To compare the denoising effect, each recommender is trained with the following approaches:
\begin{itemize}
    \item \textbf{BCE} represents the model is trained with the base pointwise binary cross-entropy loss.
    \item \textbf{BPR} \shortcite{Rendle2009BPRBP} represents the model is trained with the pairwise Bayesian Personalized Ranking loss.
    \item \textbf{T-CE} \shortcite{Wang2021DenoisingIF} removes the samples with higher loss through dynamic threshold.
    \item \textbf{R-CE} \shortcite{Wang2021DenoisingIF} is guided by a similar assumption as T-CE, while allocating lower weights to noise samples.
    \item \textbf{RGCF} \shortcite{RGCF} discard noisy edges according to the structure representation cosine similarity and enhance the diversity with graph self-supervised learning.  
    \item \textbf{DCF} \shortcite{he2024double} gradually relabel noisy samples to address the scarcity issue when denoising.
\end{itemize}

\subsubsection{Implementation Details.}
\label{sec:ImpleDetails}
All methods are trained with a batch size of 1024 and a learning rate of 0.005 with Adam optimizer for up to 200 epochs.
We adopt the early stopping during training.
We adopt the RecBole implementation for all backbone models.
Hyper-parameters are selected based on the origin setting.
We did a grid search on the following hyper-parameters to find the optimal result for LLMHD, including
$\alpha$ and $\varepsilon_{l}^{max}$, which are explored within \{3k, 5k, 10k, 30k, 50k\} and \{0.01, 0.03, 0.05, 0.1, 0.2\} respectively.
The hard sample proportion $\varepsilon_{v}$ is selected from \{0.1, 0.3, $\cdots$, 0.9\}.
All thresholds are fixed as follows, $\varepsilon_{pos}^{max}$ =8, $\varepsilon_{pos}^{min}$ = 6, $\varepsilon_{neg}^{max}$ = 4 and $\varepsilon_{neg}^{min}$ = 2 for pointwise hard sample identification.
The pairwise $\varepsilon_{pair}^{max}$ and $\varepsilon_{pair}^{min}$ are fixed as 7 and 3 respectively.
The confidence threshold $\varepsilon_{\gamma}$ to update user preference is explored within \{3,5,7,9\}.

\subsection{Performance Comparison (RQ1) \label{sec:performance}}
We evaluated the effectiveness of our proposed method in both pointwise $\text{LLMHD}_{\text{BCE}}$ and pairwise $\text{LLMHD}_{\text{BPR}}$ setting.
The performance against other denoising recommendation strategies is shown in Table ~\ref{table:performance}.
Notably, LLMHD significantly enhances the performance of all backbone models trained with normal BCE or BPR on all datasets, demonstrating its superior denoising capability.
Moreover, LLMHD consistently outperforms other advanced denoising methods across the majority of datasets and backbones.
We attribute this improvement to the extended hard sample identification, where baselines like T-CE and RGCF lack the capability, rendering them less effective in comparison.
We also observed that for most datasets and backbones, RGCF and DCF are inferior to us alone.
In contrast, other denoise baselines like T-CE and R-CE perform worse than them.
This can be explained by the fact that both RGCF and DCF are designed to insert or preserve high-confidence interactions, a feature not inherent to T-CE and R-CE.
Since LLMHD also focuses on retaining more interactions (i.e., hard samples), we posit that maintaining a more extensive set of samples is beneficial in enhancing performance.

\subsection{Noise Robustness (RQ2)}
\begin{figure}[t]
    \centering
    \includegraphics[width=\linewidth]{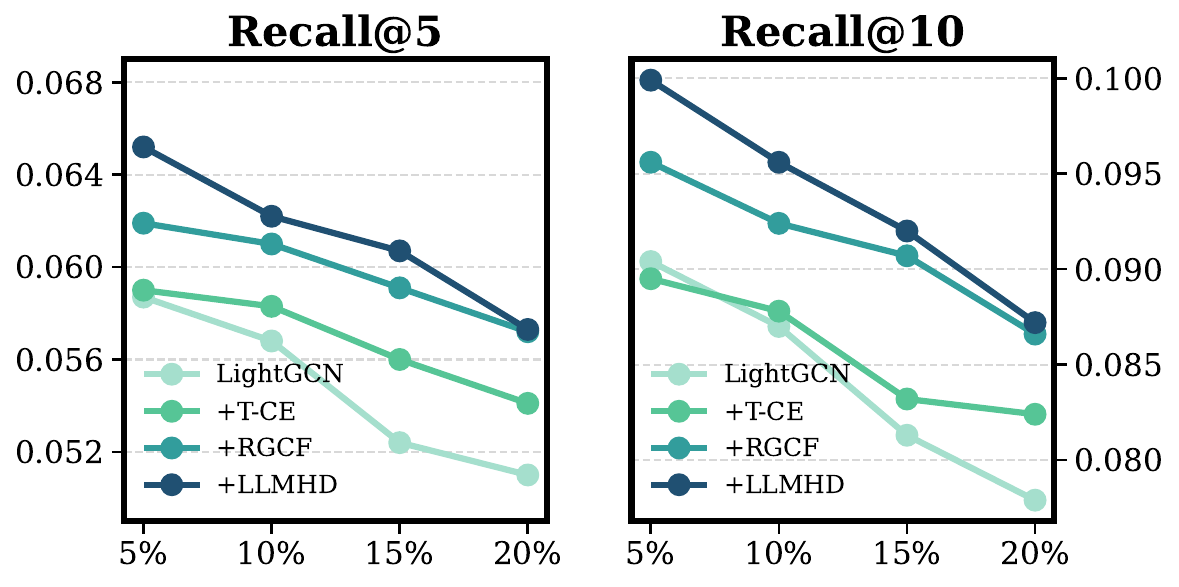}
    \caption{Performance comparison of denoise training with random noises in Amazon-books.}
    \label{fig:NoiseRobust}
\end{figure}

We conduct random noise training to evaluate the robustness of the noise resistance capability of LLMHD, comparing it with the most competitive RGCF and the classical approach T-CE.
Following previous work \shortcite{Wu2020SelfsupervisedGL}, the proportion of noise injected into the training set spanned from 5\% to 20\%, while keeping the samples in the testing set unchanged.
We report the result in Figure \ref{fig:NoiseRobust}.
The result shows that:
(1) As the noise ratio increases, we observe a consistent downward trend in the performance across all backbone models and denoising strategies.
This decline can be caused by the rising noise level leading to the increasing data corruption, which complicates discerning genuine user preferences.
(2) LLMHD outperforms the backbone model and other denoise approaches in all noise ratios.
This emphasizes LLMHD's promising noise resistance, attributed to the correctly identified hard and noisy samples.
(3) we also observed that RGCF shows sub-optimal results and slower performance degradation with increasing noise.
This is probably because it employs random edge augmentation, which makes the model adapt to Gaussian noise.
However, when confronted with real-world data noise, i.e., when additional noise is smaller, its performance falls short of LLMHD.

\subsection{In-depth Model Analysis (RQ3)}

\subsubsection{Ablation Study.}
\begin{table}[t]
    \setlength{\tabcolsep}{5pt}
    \small
    \begin{tabular*}{\linewidth}{l|cccc}
        \toprule
        \multirow{2}*{\textbf{Methods}} & \multicolumn{4}{c}{\textbf{Amazon-books}} \\
        ~ & \textbf{R@5} & \textbf{R@10} & \textbf{N@5} & \textbf{N@10} \\
        \midrule
        LightGCN & 0.0587 & 0.0904 & 0.0598 & 0.0704 \\
        \midrule
        + LLMHD$_{\text{LD}}$ & 0.0614 & 0.0965 & 0.0628 & 0.0743 \\
        + LLMHD$_{\text{LD} + \text{RS} + \text{LMS}}$ & 0.0623 & 0.0970 & 0.0635 & 0.0749 \\
        + LLMHD$_{\text{LD} + \text{VS} + \text{LMS}}$ & 0.0638 & 0.0986 & 0.0655 & 0.0767 \\
        + LLMHD$_{\text{LD} + \text{VS}+\text{LMS} + \text{PU}}$ & \textbf{0.0652} & \textbf{0.0999} & \textbf{0.0665} & \textbf{0.0771} \\
        \bottomrule
    \end{tabular*}
    \caption{The effect of each components in LLMHD$_{\text{BPR}}$ with the LightGCN on Amazon-books dataset.}
    \label{tab:ablation}
\end{table}

We conduct experiments to assess each module in LLMHD, including Variance-based Sample Pruning, LLM-based Sample Scoring, and Iterative Preference Updating, the result is shown in Table \ref{tab:ablation}.
(1) We investigate whether including Variance-based Sample Pruning (VS) enables an effective hard sample candidate selection.
Specifically, we compare it with Random Selection (RS) and select the same amount of candidates as the VS.
According to the result, converting the VS to RS leads to a performance drop in all metrics.
This reveals the superiority of the Variance-based Sample Pruning in selecting hard sample candidates.
(2) We discover whether LLM-identified hard samples enhance the recommendation performance.
The comparison is made between the backbone that only adopts a Loss-based Denoise Module (LD) and the one that includes LLM-based Sample Scoring (LMS).
Significant improvement is demonstrated after using LLMs to detect hard samples, indicating the advancement of LLM in hard sample identification.
(3) We explore the influence of adopting Iterative Preference Updating (PU).
Compared with discarding the preference updating, the performance of adopting it increases, demonstrating the effectiveness of Preference Updating in understanding genuine user preference.

\subsubsection{Effect of Hyper-parameters.}
\begin{figure}[t]
    \centering
    \includegraphics[width=\linewidth]{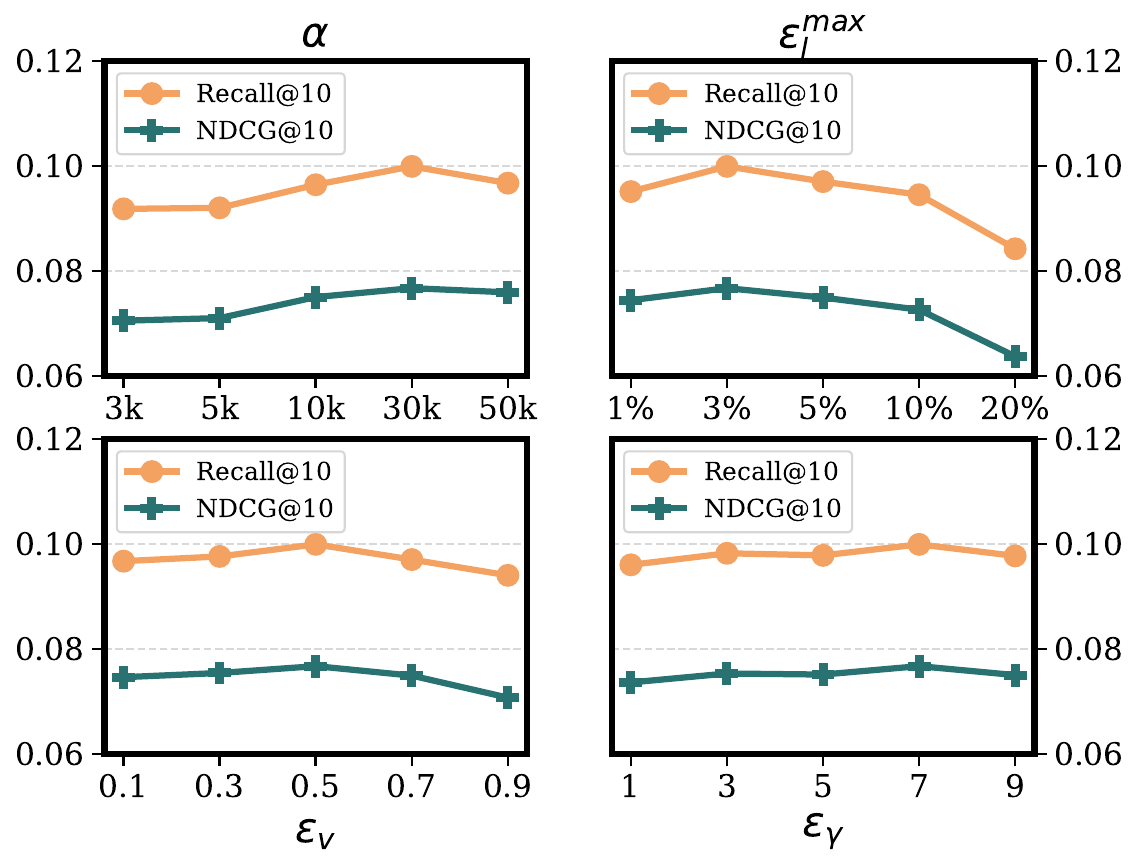}
    \caption{Hyper-parameter analysis in $\text{LLMHD}_{\text{BPR}}$ with LightGCN backbone on the Amazon-books.}
    \label{fig:Hyperparams}
\end{figure}

For a more elaborate analysis, we adjust the hyper-parameters within the range described in the \nameref{sec:ImpleDetails} section.
The results are shown in Figure \ref{fig:Hyperparams}.
From our observations:
(1) Growth Rate $\alpha$: A moderate increase in $\alpha$ enhances performance, as it retains more samples per iteration, mitigating data scarcity during training. However, excessively high values degrade performance.
(2) Max Noise Scale $\varepsilon_{l}^{max}$: Elevating $\varepsilon_{l}^{max}$ initially improves LLMHD by filtering out more noise, but an overly high setting results in excessive sample loss, hampering the learning of user preferences.
(3) Hard Sample Candidate Proportion $\varepsilon_{v}$: Increasing $\varepsilon_{v}$ presents more hard sample candidates, boosting performance. But setting it too high may confuse noisy samples for hard ones, lowering overall effectiveness.
(4) Confidence Threshold $\varepsilon_{\gamma}$: Gradually raising $\varepsilon_{\gamma}$ initially benefits the model by promoting item selection for preference update.
However, a high confidence restricts item discovery and a low confidence finds incorrect items, both diminishing user preference understanding.

\section{Conclusion}

In this work, we introduced the Large Language Model Enhanced Hard Sample Denoising (LLMHD) framework to address the challenge of distinguishing hard samples from noise samples for recommender systems.
By utilizing an LLM-based scorer to evaluate semantic consistency between users and items and assessing sample hardness according to its compatibility with training objectives, we can differentiate hard samples from noise samples.
We further introduce a variance-based sample pruning strategy to effectively select candidates.
In addition, the iterative preference update refines user preference and mitigates biases introduced by false-positive interactions.
Extensive experiments on real-world datasets and recommenders demonstrated the effectiveness of LLMHD in improving recommendation quality. 

\bibliography{main}

\section{Supplementary Materials}

\subsection{Details of Dataset}
In this section, we offer details of the preprocessed dataset adopted in the experiment.
We take the datasets in RLMRec \cite{Ren2023RepresentationLW}, in which each item contains a corresponding item text profile.
Therefore, we follow the preprocessing setting in the RLMRec.
Specifically, interactions with ratings below 3 for both the Amazon-books and Yelp data are filtered out.
No rating-based filtering is adopted in Steam.
K-core filtering is also performed and split into training, validation, and test sets using a 3:1:1 ratio.
The statistics of datasets preprocessed following RLMRec are shown in Table \ref{tab:dataset}.

\begin{table}[h]
    \centering
    \setlength{\tabcolsep}{3.5pt}
    \small
    \begin{tabular*}{\linewidth}{c|cccc}
        \toprule
        Datasets & \# Users & \# Items & \# Interactions & \# Sparsity \\
        \midrule
        Amazon-books & 11,000 & 9,332 & 120,464 & 99.88\% \\
        Yelp & 11,091  & 11,010 & 166,620 & 99.86\% \\
        Steam & 23,310  & 5,237 & 316,190 & 99.74\% \\
        \bottomrule
    \end{tabular*}
    \caption{Statistics of preprocessed datasets.}
    \label{tab:dataset}
\end{table}

However, previous works adopted the rating score to label noise and clean data.
For example, T-CE regards a rating score below 3 as a false-positive interaction.
As a result, the dataset filtered with ratings in RLMRec is regarded to contain less noisy interactions.
To compare the denoising ability of different methods, we add 5\% noisy interactions to the training set.
These noisy interactions are selected from the interactions that are rated below 3.
Experiments are then conducted on these noise-inserted datasets.

\subsection{Details of API Token Cost}

\begin{table}[t]
    \centering
    \small
    \setlength{\tabcolsep}{9pt}
    \begin{tabular*}{\linewidth}{c|cccc}
        \toprule
        Datasets & Template & $\varepsilon_{l}^{max}$ & $\varepsilon_{v}$ & \# Token  \\
        \midrule
        
        \multirow{6}*{Amazon-books} & \multirow{1}*{$T_{sum}$} & - & - & 16m \\
        ~ & \multirow{1}*{$T_{FN}$} & - & - & 6m \\
        ~ & \multirow{1}*{$T_{FP}$} & - & - & 3m \\
        \cline{2-5}
        ~ & \multirow{3}*{$T_{score}$} & 0.05 & 0.5 & 11m  \\
         ~ & ~ & 0.10 & 0.5 & 21m  \\
        ~ & ~ & 0.05 & 0.3 & 7m \\
        
        \midrule
        \multirow{6}*{Yelp} & \multirow{1}*{$T_{sum}$} & - & - & 19m \\
        ~ & \multirow{1}*{$T_{FN}$} & - & - & 8m \\
        ~ & \multirow{1}*{$T_{FP}$} & - & - & 3m \\
        \cline{2-5}
        ~ & \multirow{3}*{$T_{score}$} & 0.05 & 0.5 & 16m  \\
         ~ & ~ & 0.10 & 0.5 & 30m  \\
        ~ & ~ & 0.05 & 0.3 & 10m  \\
        
        \midrule
        \multirow{6}*{Steam} & \multirow{1}*{$T_{sum}$} & - & - & 40m \\
        ~ & \multirow{1}*{$T_{FN}$} & - & - & 10m \\
        ~ & \multirow{1}*{$T_{FP}$} & - & - & 4m \\
        \cline{2-5}
        ~ & \multirow{3}*{$T_{score}$} & 0.05 & 0.5 & 35m  \\
         ~ & ~ & 0.10 & 0.5 & 62m  \\
        ~ & ~ & 0.05 & 0.3 & 20m  \\
        
        \bottomrule
    \end{tabular*}
    \caption{OpenAI API token number.}
    \label{tab:apicost}
\end{table}

Since we adopt the GPT-3.5-turbo as the LLM in the LLMHD, here we provide the token number for training with LLMHD in Table \ref{tab:apicost}.
The total token number of $T_{score}$ in LLMHD is highly dependent on two aspects:
the number of interactions in the dataset, the value of the maximum noise scale $\varepsilon_{l}^{max}$, and the hard sample proportion $\varepsilon_{v}$. 
Whears, the $T_{sum}$ is not correlated to the hyper-parameters.
We also report the token number for $T_{FN}$, and $T_{FP}$ during training when the confidence threshold is set to $\varepsilon_{\gamma}=3$.

\subsection{Details of Loss Value and Prediction Score Figure}
We provide the details of plotting the Figure \ref{fig:Introduction}.
For the noise samples, we follow the settings of \cite{Wang2021DenoisingIF}, taking the interactions that rate below 3 as the false-positive noise.
By flipping labels of ground truth records in the test set, we can obtain a set of false-negative interactions that are positively labeled but unobserved during the negative sampling process.
Samples in which the positive item is false-positive and the negative item is false-negative are considered noisy samples.
We then identify hard and easy samples according to the setting in \cite{Ding2020Simplify}.
For each positive interaction, $D$ negative items are sampled. 
Then the negative item with the highest prediction score of $\hat{y}_{u,i}$ is adopted as the negative instance during training.
Thus, when the $D$ gets higher, the sample becomes harder.
According to this setting, we collect the prediction score and loss value results when $D=1$ and that of hard samples when $D=3$.

\subsection{Details of Prompt Template}

\begin{figure*}[!t]
  \centering
  \includegraphics[width=\linewidth]{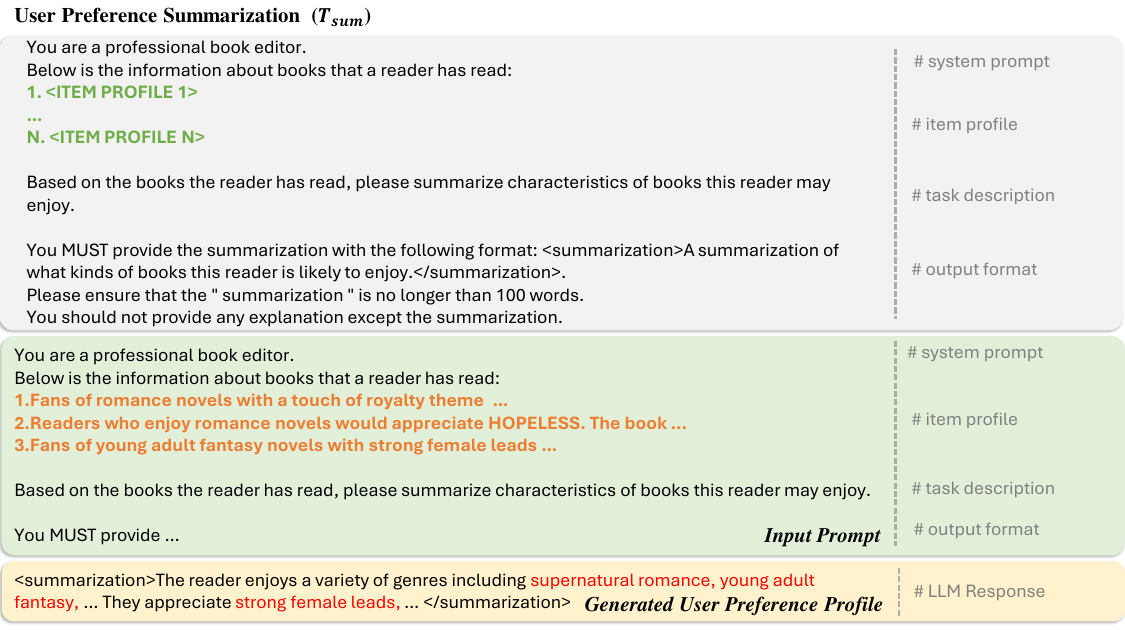}
  \caption{
        Example of user preference summarization process on Amazon-books dataset.
    }
  \label{fig:UPS}
\end{figure*}

\begin{figure*}[!t]
  \centering
  \includegraphics[width=\linewidth]{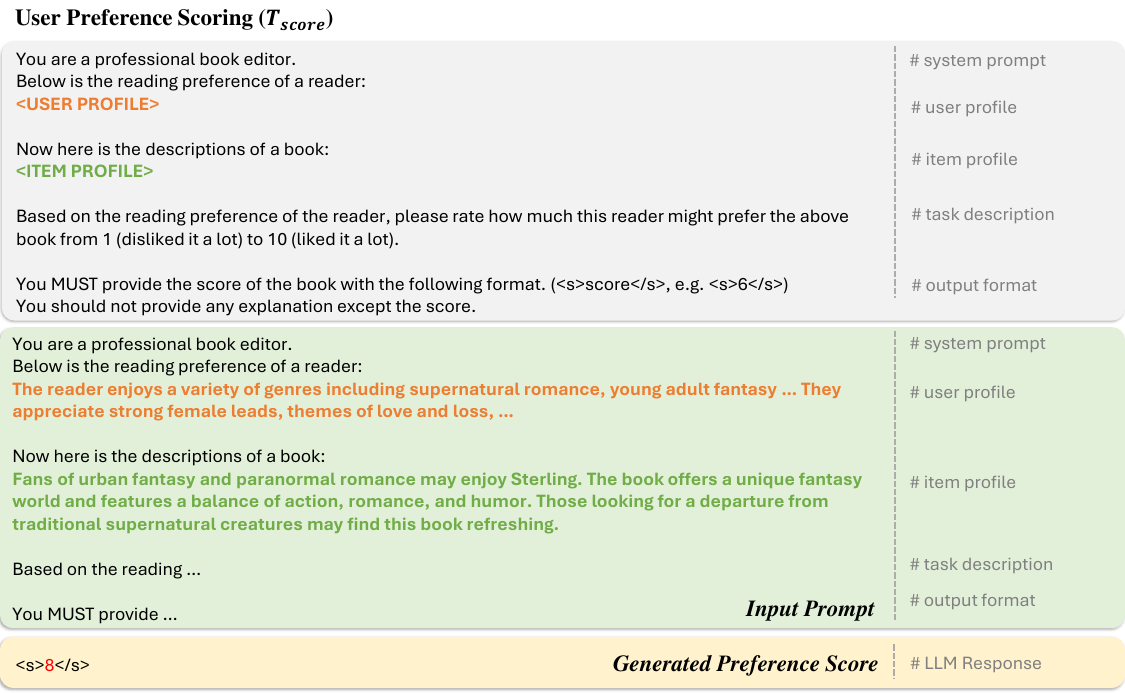}
  \caption{
        Example of user preference scoring process on Amazon-books dataset.
    }
  \label{fig:LSS}
  \vspace{-15pt}
\end{figure*}

\begin{figure*}[!t]
  \centering
  \includegraphics[width=\linewidth]{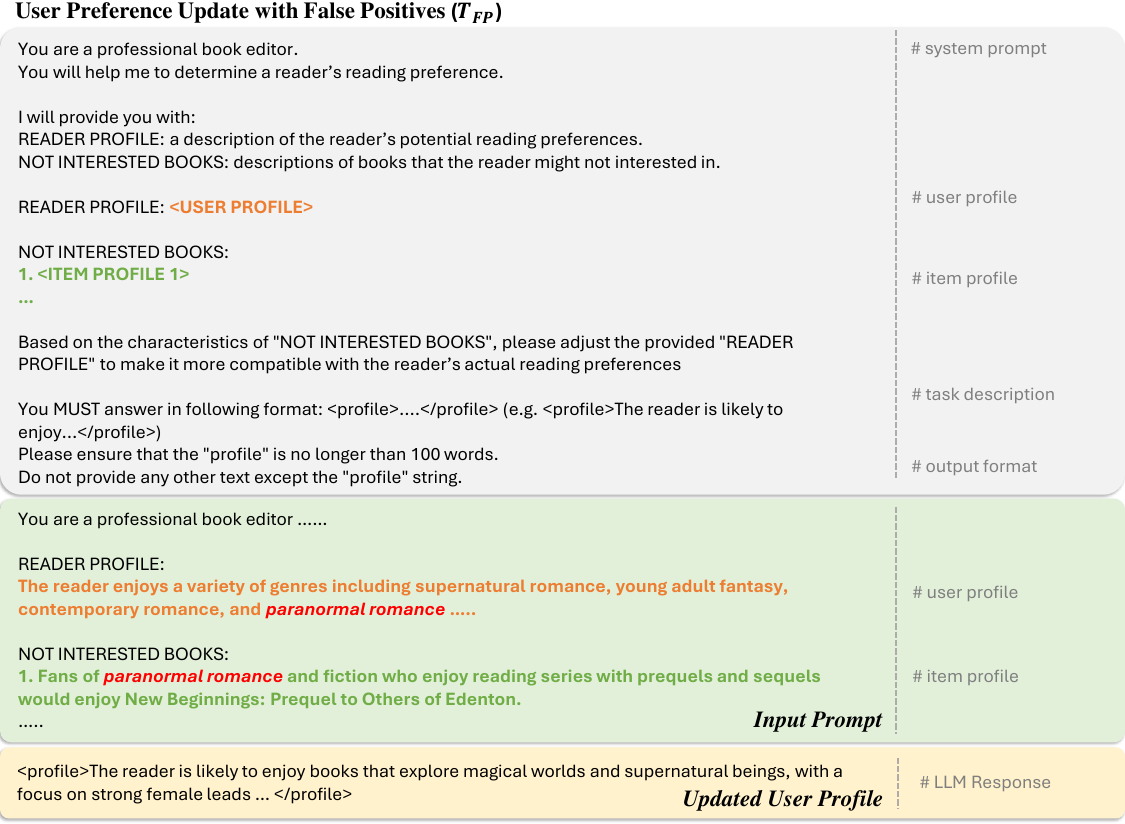}
  \vspace{-20pt}
  \caption{
       Example of update user preference with False-positive item.
    }
  \label{fig:UPUFP}
  \vspace{-5pt}
\end{figure*}

\begin{figure*}[!t]
  \centering
  \includegraphics[width=\linewidth]{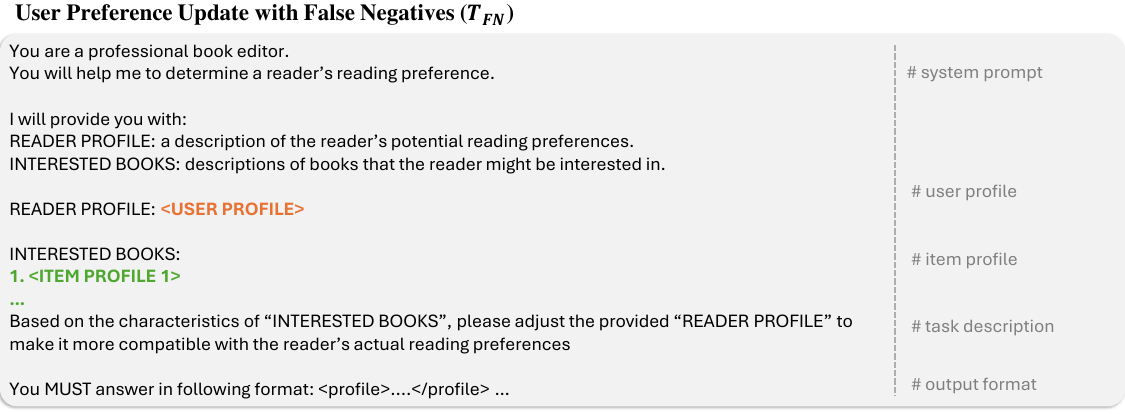}
  \vspace{-20pt}
  \caption{
        Example of update user preference with false-negative item.
    }
  \label{fig:UPUFN}
  \vspace{-15pt}
\end{figure*}

In this section, we offer comprehensive information on the templates utilized in the LLMHD.
Real examples from the Amazon-book dataset are used as a showcase.
The templates used in the Yelp and Steam datasets are with minor differences in the instructions provided to represent different data.

\subsubsection{Example of User Preference Summarization.\label{sec:Appendixups}}
Figure \ref{fig:UPS} cases an example of user preference profile generation, specifically for the Amazon-book dataset.
The instruction provided to the language model for all users remains consistent, directing the LLMs to summarize the characteristics of books that would appeal to the user.
The input item profiles consist of the book title and a corresponding item description from the dataset.
To facilitate parsing the output, we enforce the output format to the XML.
The generated result demonstrates that the LLMs, in this case ChatGPT, capture the common features from the interacted items.

\subsubsection{Example of LLM Sample Scoring.}
Figure \ref{fig:LSS} illustrates the process of scoring user preference for an item with LLMs.
In order to achieve the user preference for a specific item, the prompt template incorporates the user preference profile generated in the User preference Profile generation \ref{sec:Appendixups} and the item profile in the dataset.
By utilizing both information, LLMs are empowered to infer the user's preference.
In the presented example, leveraging the user preference profile and item profile, the large language model assesses the preference for the book accurately.
In addition, to maintain consistency, we ask the model to generate scores from 1 to 10, and correlated descriptions about the meaning of different scores are also provided.
This enables the model to generate scores in a specific range, making subsequent judgments of hard samples easier.

\subsubsection{Example of User Preference Update.}

Figure \ref{fig:UPUFP} shows the overall process of updating user preference with False-positive items.
With the provided item profile and user profile, the LLM successfully refined the provided user preference profile by removing correlated user-dislike item characteristics.
This demonstrates the promising ability of LLMs to understand actual user preferences by providing user-like or disliked items.
A similar effect is also achieved by updating user preference with False-negative items, where the prompt template is shown in Figure \ref{fig:UPUFN}.

\end{document}